\documentclass[hidelinks,12pt]{article}

\usepackage{arxiv}

\usepackage[utf8]{inputenc} 
\usepackage[T1]{fontenc}    
\usepackage{hyperref}       
\usepackage{url}            
\usepackage{booktabs}       
\usepackage{amsfonts}       
\usepackage{nicefrac}       
\usepackage{microtype}      
\usepackage{lipsum}
\usepackage{graphicx}
\graphicspath{ {./images/} }
\usepackage{siunitx}
\usepackage[onehalfspacing]{setspace}

\title{Comparison of deep-learning data fusion strategies in mandibular osteoradionecrosis prediction modelling using clinical variables and radiation dose distribution volumes}

\author{
 Laia Humbert-Vidan, PhD \\
  Department of Medical Physics\\
  Guy's and St Thomas' NHS Foundation Trust\\
  School of Cancer and Pharmaceutical Sciences\\
  King’s College London\\
  London, UK\\
  \texttt{laia.humbert-vidan@kcl.ac.uk} \\
  \And
  Vinod Patel, PhD \\
  Department of Oral Surgery\\
  Guy’s and St Thomas’ NHS Foundation Trust\\
  London, UK \\
  \And
  Andrew P King\thanks{Joint last authors.} \\
  School of Biomedical Engineering \& Imaging Sciences\\
  King's College London\\
  London, UK \\
  \And
  Teresa Guerrero Urbano\footnotemark[1] \\
  Department of Clinical Oncology \\
  Guy's and St Thomas' NHS Foundation Trust \\
  School of Cancer and Pharmaceutical Sciences\\
  King’s College London\\
  London, UK\\
}

\begin{document}
\maketitle
\begin{abstract}
\textit{Purpose.} Normal tissue complication probability (NTCP) modelling is rapidly embracing deep learning (DL) methods as the need to include spatial dose information is acknowledged. Finding the most appropriate way of combining radiation dose distribution images and clinical data involves technical challenges and requires domain knowledge. We propose different data fusion strategies that we hope will serve as a starting point for future DL NTCP studies.
\textit{Materials and methods.} Early, joint and late DL multi-modality fusion strategies were compared using clinical variables and mandibular radiation dose distribution volumes. The discriminative performance of the multi-modality models was compared to that of single-modality models: a random forest trained on non-image data (clinical, demographic and dose-volume metrics) and a 3D DenseNet-40 trained on image data (radiation dose distribution volumes of the mandible). All the experiments were conducted on a control-case matched cohort of 92 ORN cases and 92 controls from a single institution.
\textit{Results.} The highest ROC AUC score was obtained with the late fusion model (0.70), but no statistically significant differences in discrimination performance were observed between strategies. While late fusion was the least technically complex strategy, its design did not model the inter-modality interactions that are required for NTCP modelling. Joint fusion involved the most complex design but resulted in a single network training process which included intra- and inter-modality interactions in its model parameter optimisation.
\textit{Conclusions.} To our knowledge, this is the first study that compares different strategies for including image data into DL NTCP models in combination with lower dimensional data such as clinical variables. The discrimination performance of such multi-modality NTCP models and the choice of fusion strategy will depend on the distribution and quality of both types of data. We encourage future DL NTCP studies to report on different fusion strategies to better justify their choice of DL pipeline.
\end{abstract}


\section{Introduction}
Radiation-induced toxicities arise from the combination of radiation dose, clinical and demographic information as risk factors. Conventional normal tissue complication probability (NTCP) models include one-dimensional metrics (e.g. mean dose) calculated from the dose-volume histogram (DVH) as the dosimetric information. DVH metrics are then typically combined with clinical and demographic factors to find statistical associations with toxicity outcome.

A DVH is a simplified representation of the three-dimensional radiation dose distribution in an anatomical structure such as an organ at risk (OAR) and does not include spatial dosimetric information. As a result, DVH metrics (e.g. mean dose) might not be representative of the dose distribution, especially if the radiation dose is not uniformly distributed across the structure. How the radiation dose is spatially distributed within the OAR is clinically relevant as anatomical heterogeneities within the structure will result in organ sub-regions with different radiation responses \cite{Hopewell2000,Marks2010}. Spatial dose associations with radiation-induced toxicities in head and neck cancer (HNC) have been studied. Therefore, including such spatial dose information into toxicity prediction models has the potential to result in similar or even improved prediction accuracies to conventional DVH-based models \cite{Dean2016,Monti2017,Gabrys2018,Ebert2021}.

Deep learning (DL) has proved to be an effective method for working directly with the three dimensional dose maps and extracting the relevant spatial image features that contribute most to the toxicity prediction \cite{Ibragimov2018,Men2019,Humbert-Vidan2022a,Reber2023,ElNaqa2022,Appelt2022}. Existing radiation toxicity prediction studies \cite{Ibragimov2018} have combined dose maps with clinical and demographic factors but the impact of the architectural design choices in data fusion DL pipelines on the prediction performance has yet to be investigated.

DL multimodality data fusion strategies have been classed as \textit{early or feature level} fusion, \textit{joint or intermediate} fusion and \textit{late or decision level} fusion, based on when the fusion of the two data types takes place \cite{Huang2020,Ramachandram2017}. In early fusion, the inputs from different modalities, some of which may be features extracted by a machine-learning (ML) algorithm, are combined into a single vector that is then fed into a single ML model. In joint fusion, at least one of the modalities makes use if features learned using a feature extraction neural network. The combined features are input into a final neural network, the loss of which is backpropagated to the first feature extraction neural network model(s). Finally, in late fusion, the final predictions from multiple models are combined to make a final decision. In both early and joint fusion strategies, the features may be original or extracted with a ML (e.g. image features extracted with a convolutional neural network, CNN). Both early and joint fusion strategies can model the interactions between features from different modalities. Joint fusion is thought to result in better feature representations due to the backpropagation of the combined model loss to the feature extraction neural networks during training. However, joint fusion can result in a more complex network design than early or late fusion.

Mandibular osteoradionecrosis (ORN) has a multifactorial aetiology where radiation dose plays an important role along with other clinical and demographic risk factors \cite{Frankart2021,Patel2022,DeFelice2020}. Several studies \cite{Mohamed2017,Moon2017,Aarup-Kristensen2019,Moring2022} have investigated the associations between these factors and the incidence of ORN and a DVH-based NTCP model was recently developed by van Dijk et al. \cite{vanDijk2021}. In this study we compare different multimodality data fusion strategies in the context of prediction of mandibular ORN in HNC patients treated with radiotherapy. We propose DL and ML pipelines for the different fusion strategies that we hope will be easily applicable to other contexts within the NTCP modelling field.

\section{Materials and methods}
\label{sec:headings}
\subsection{Clinical variables}
Clinical and demographic data (Table 1) were collected for a total of 92 mandibular ORN subjects treated with intensity-modulated radiation therapy at our institution between 2011 and 2022. During the time span considered, from a total of 1721 HNC patients radically treated, a total of 142 patients (8.3\%) were diagnosed with ORN, 50 of which were excluded because of unavailable RT dose and/or RT plan files (18), ORN region outside of the mandible (15), palliative or low prescribed dose (8), previous irradiation in the HN region (6) or two primary tumour sites (3). A control group of 92 subjects was selected with a 1:1 control-case matching approach based on primary tumour site and treatment year. Primary tumour site groups considered included oral cavity, oropharynx, paranasal sinus/nasopharynx, larynx/hypopharynx, salivary glands and unknown primary. No minimum follow-up time threshold was applied for the control group. Thus, the average follow-up time for the control group was 49.9 months (range 5.2-92.0) while the median time from the end of RT to diagnosis of ORN was 12.1 months (IQR 20.3). 

The Chi-squared statistical test (Mann Whitney U test for the continuous variable ‘Age’) was used to evaluate the difference in the distribution of the categorical variables between the ORN group and the control group. Categorical variables were dichotomised, and the continuous variable ‘Age’ was normalized within the range 0-1.

Patients who develop ORN after their RT course at our centre are treated and closely monitored by a specialist oral surgical team in a dedicated clinic. The Notani \cite{Notani2002} ORN staging system was used; however, for the purpose of binary classification in the experiments presented, toxicity outcomes were dichotomised and any grade of ORN was considered as an event.

\begin{table}[h]
    \caption{Distribution of clinical and demographic variables with the corresponding p-value for the comparison between groups. Smoking status and alcohol consumption were positive if reported as such within two months of the radiotherapy start.}
    \centering
    \begin{tabular}{lccc}
        \toprule
         &  ORN &  Control  & p-value \\
         \midrule
         Age (median (IQR))& 62 (13) & 61 (15) & 0.455\\
         Gender (male)&  66 (72\%)&72 (78\%)  & 0.395\\
         Smoking (current)&  47 (51\%)&  21 (23\%) & <0.001\\
         Alcohol (current)&  71 (77\%)& 63 (69\%) & 0.246\\
         Pre-RT extractions& 55 (60\%) & 50 (54\%) & 0.551\\
         Pre-RT surgery (PORT)&  35 (38\%)& 35 (38\%) & 1.000\\
         Chemotherapy &  59 (64\%)& 57 (62\%) & 0.879\\         
         \bottomrule
    \end{tabular}
    \label{tab:Table1}
\end{table}

\subsection{Radiation dose distribution maps}
The mandible was manually contoured in all patients on the RT planning CT by a single user to ensure consistency. The RT planned dose distribution was masked by the mandible structure to obtain the mandible dose distribution map. Image processing included the following steps: image resampling, common space registration with the reference image being that of the subject with the largest mandible volume, image normalization to a voxel value range between 0 and 1, image cropping to exclude slices not containing mandible contours (if necessary, smaller mandible volumes were padded with empty slices to achieve a consistent number of slices, z=56, across the cohort) and image resizing to 64x64x56.

\subsection{Single modality models}

\subsubsection{Random forest}
A random forest (RF) binary (ORN vs. no ORN) classifier \cite{Breiman2001} was trained on the clinical and demographic variables and implemented using the \textit{sklearn.ensemble.RandomForestClassifier} Python module. Stratified 5-fold nested cross-validation was used, with embedded model hyperparameter optimisation implemented with the \textit{GridSearchCV} Python module. For the grid search CV procedure, the following hyperparameters were considered: bootstrap (True, False), maximum depth (1, 2, 10, 20, None), maximum features (auto, sqrt), minimum samples per leaf (1, 2, 4), minimum samples per split (2, 5, 10) and number of estimators (200, 500, 1000, 1300, 1700, 2000).

\subsubsection{3D DenseNet-40 CNN}
A 3D DenseNet-40 (DN40) CNN was trained for binary classification with the mandible dose distribution maps as the single input and implemented using the Medical Open Network for Artificial Intelligence (MONAI) (https://monai.io/) Pytorch-based framework. The 3D DN40 is a shallower version of the more commonly used DenseNet configurations \cite{Huang2017}; it has a total of three dense blocks, each with 12 dense layers, interleaved with two transition blocks. The convolutional, pooling, batch normalisation and dropout operations included in the dense and transition blocks are all three-dimensional. After the third dense block, a final 3D average pooling layer reduces the output to one dimension and this output is flattened before it is passed through the final fully connected layer and a Softmax layer that provides the final classification probabilities for the binary outcome. 

The data were split into training, validation and test sets following a stratified 5-fold nested CV approach as for the RF model. The Adam optimisation algorithm and the categorical cross entropy loss function (\textit{torch.nn.CrossEntropyLoss}) were used. A hyperparameter grid search was performed which included the following hyperparameters and values: dropout 0.6, 0.8; learning rate 0.01, 0.001, 0.0001; batch size 10, 16; weight decay 0.01, 0.001, 0001; epochs 50, 100, 300. Small 3D random rotation (\SIrange{-0.1}{0.1}{\radian}) and zoom (0.8 to 1.2) data augmentations were applied to the dose maps of the training set.

\subsection{Multimodality fusion strategies}
Three fusion strategies (Figure 1) were explored and compared for combining the dose distribution maps and clinical variables into a DL/ML pipeline for the binary classification of ORN vs. no ORN.

\begin{figure}[htbp]
    \centering
    {\includegraphics[width=0.9\linewidth]{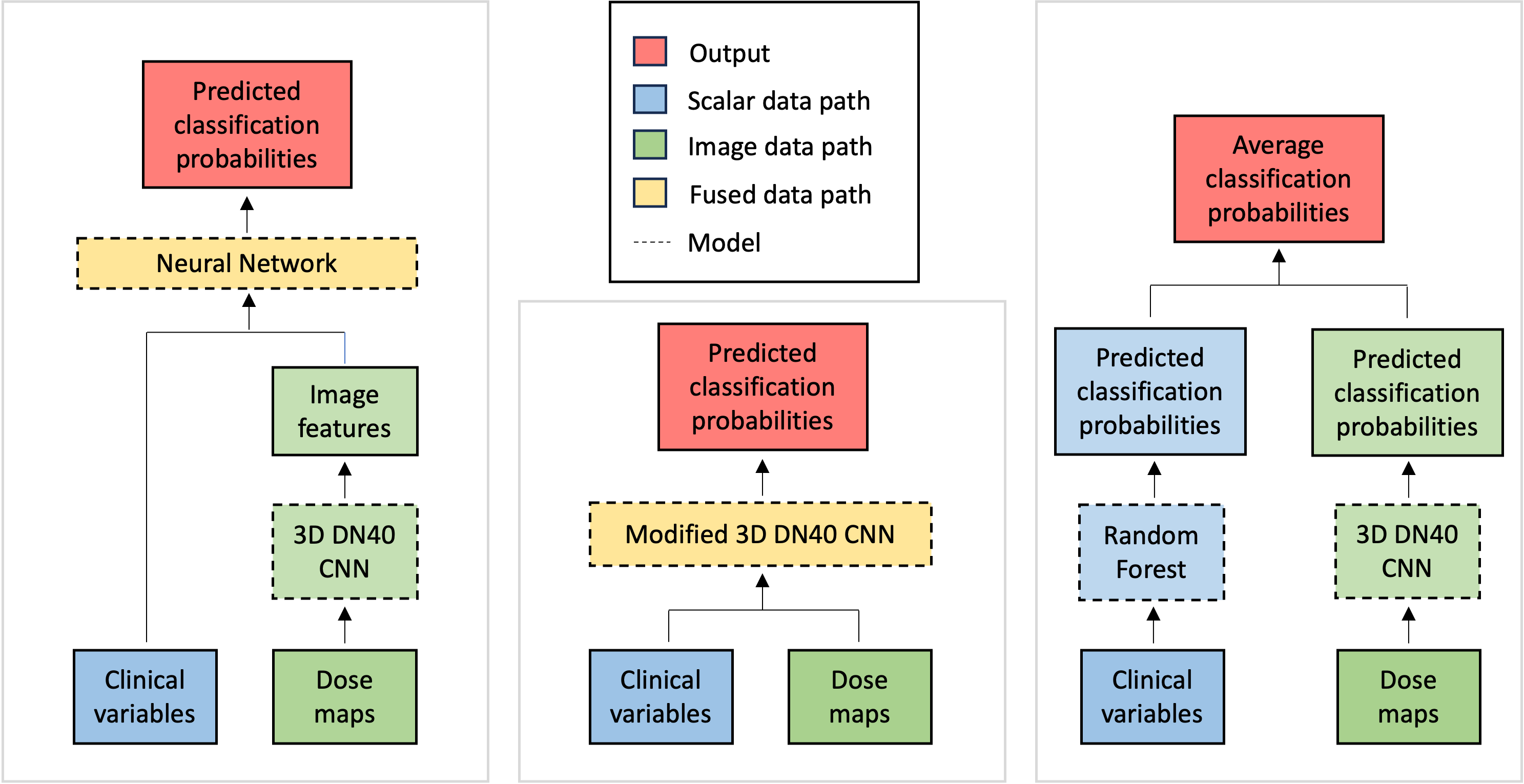}}
    {\caption{Schematics of the early (left), joint (middle) and late (right) multimodality data fusion strategies implemented.}}
    \label{fig:fig1}
\end{figure}

\subsubsection{Early fusion}
Image (mandible dose map) features were extracted from the previously trained 3D DenseNet-40 CNN model during inference on the test dataset. The resulting tensor, consisting of 688 learned image features, was flattened and concatenated with the 7 clinical variables into one single feature vector. This combined feature vector was fed into a simple classification neural network with two fully connected layers: an input linear layer with 64 hidden neurons and an output linear layer with 64 hidden neurons and two output channels. A final softmax layer was added to obtain the final predicted probabilities for each class (ORN and no ORN) (Figure 2).

\begin{figure}[htbp]
    \centering
    {\includegraphics[width=0.9\linewidth]{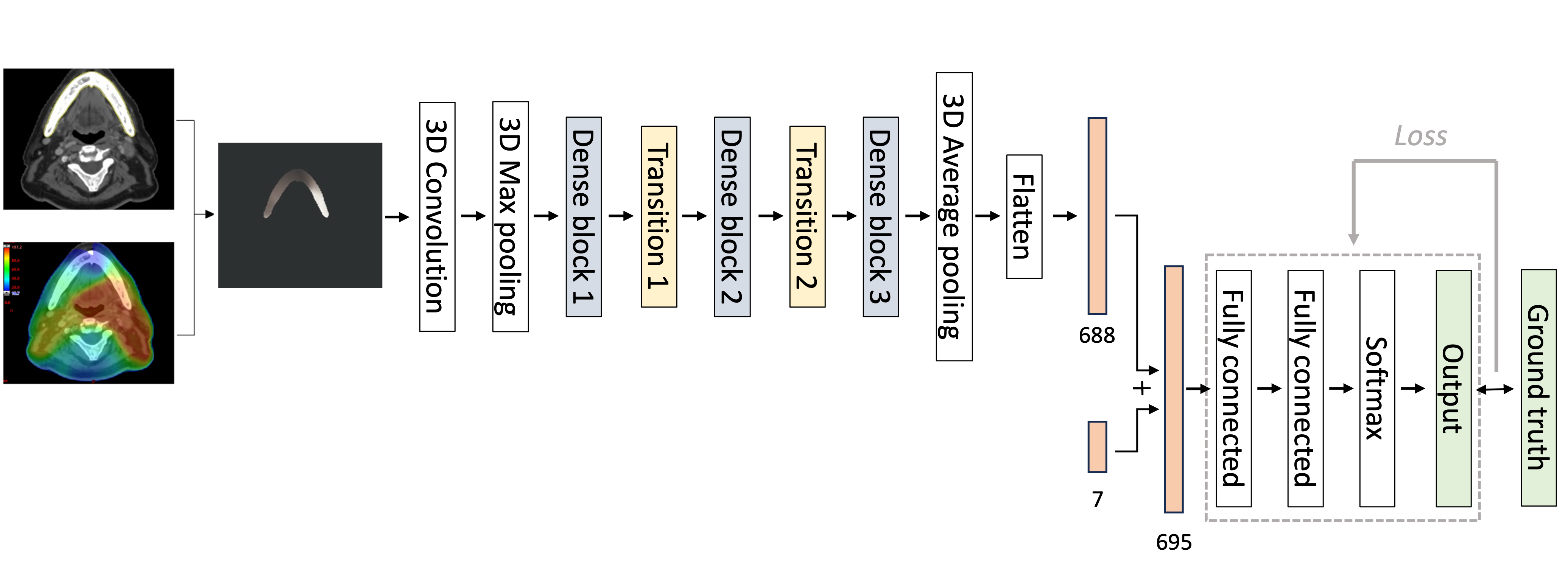}}
    {\caption{Early fusion pipeline: i) the radiation dose distribution was masked with the mandible segmentation to obtain the mandible dose map; ii) image features were extracted from the mandible dose maps with a 3D DenseNet-40 CNN; iii) the image features vector and clinical variables vector, of sizes (688,1) and (7,1) respectively, were concatenated into a single vector; iv) a classification neural network was trained on the combined feature vector (695,1) to obtain the predicted class probabilities.}}
    \label{fig:fig2}
\end{figure}

\subsubsection{Joint fusion}
A 3D DenseNet-40 classification CNN was modified with an additional concatenation layer where the clinical features (7) were concatenated with the image features (688) into one single feature vector before the final fully connected layer. Note that, unlike in the early fusion strategy, the combined feature vector in the joint fusion strategy is used to train the entire CNN, with the clinical features potentially influencing the loss value that is backpropagated during the network training process (Figure 3).

\begin{figure}[htbp]
    \centering
    {\includegraphics[width=0.9\linewidth]{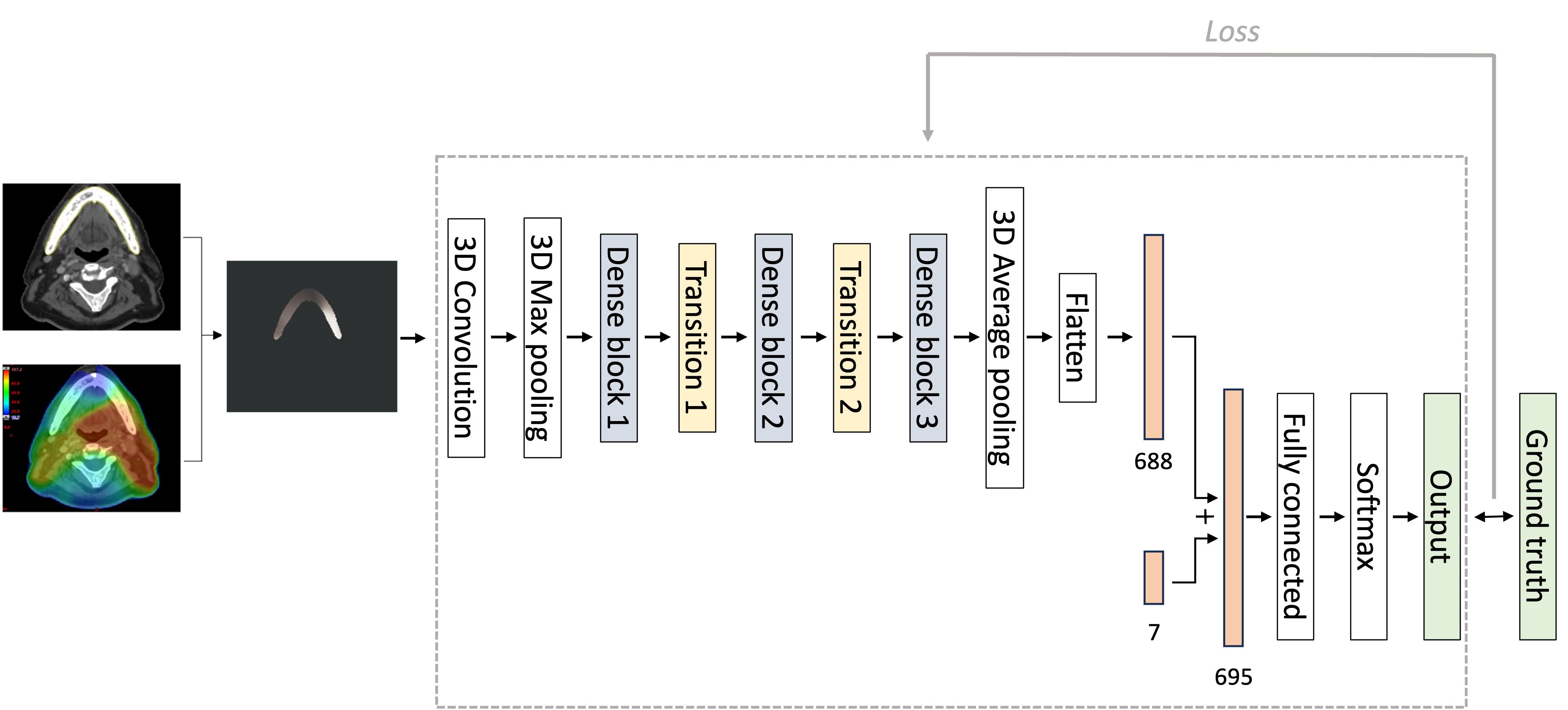}}
    {\caption{Joint fusion pipeline: the clinical variables vector was concatenated with the image features (688) extracted from the mandible dose maps before the fully connected layer of the 3D DenseNet-40 CNN. The modified 3D DenseNet-40 CNN was trained on the concatenated feature vector (695,1) to predict the final class probabilities. }}
    \label{fig:fig3}
\end{figure}

\subsubsection{Late fusion}
The predicted class probabilities were obtained separately from the RF trained on clinical variables and the 3D DenseNet-40 CNN trained on dose maps. A soft-voting ensemble approach was followed to average the two sets of classification probabilities and obtain the final class decision on a case-by-case basis for the test dataset (Figure 1).

\section{Results}

The predictive performance of the models was assessed in terms of their discriminative ability (Table 2). The ROC curves of the models (Figure 5) were compared with the DeLong nonparametric statistical test \cite{DeLong1988} using the pROC package \cite{Robin2011} with the statistical software R (https://www.R-project.org/).
The late fusion model showed the highest ROC AUC score (0.70). However, apart from the late fusion vs. RF comparison (DeLong test p-value 0.03), no statistically significant differences were observed in the discrimination performance between models.

\begin{figure}[!htbp]
    \centering
    {\includegraphics[width=0.6\linewidth]{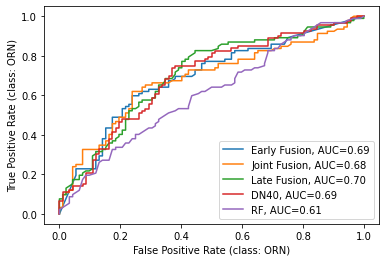}}
    {\caption{Comparison of the ROC curves for the Random Forest (RF) model trained on clinical variables, the 3D DenseNet-40 CNN (DN40) model trained on mandible dose maps and the early, joint and late multimodality fusion models. }}
    \label{fig:fig4}
\end{figure}

\begin{table}[!htbp]
    \caption{Models discrimination performance.}
    \centering
    \begin{tabular}{lccccc}
        \toprule
         &  RF&  DN40&  Early Fusion&  Joint Fusion& Late Fusion\\
         \midrule
         Accuracy&  0.59&  0.67& 0.67 & 0.69 &0.67 \\
         Sensitivity&  0.73& 0.71 &0.61  &0.65  &0.73 \\
         Specificity&  0.45& 0.63 &0.73  &0.72  &0.61 \\
         Precision& 0.57 & 0.66 &0.69  &0.70  &0.65 \\
         F1 score& 0.64 & 0.68 & 0.65 & 0.67 & 0.69\\
         \bottomrule         
    \end{tabular}
    \label{tab:Table2}
\end{table}

\begin{table}[!htbp]
    \caption{DeLong ROC AUC comparison test results}
    \centering
    \begin{tabular}{lcccc}
        \toprule
        DeLong p-value & RF & DN40 & Early Fusion & Joint Fusion\\
        \midrule
        DN40 vs. &0.09  &  &  & \\
        Early Fusion vs.& 0.08 & 0.99 &  & \\
        Joint Fusion vs. & 0.09 & 0.72 & 0.86 & \\
        Late Fusion vs. & 0.03 & 0.36 & 0.78 & 0.44\\
        \bottomrule
    \end{tabular}

    \label{tab:Table3}
\end{table}

\section{Discussion}

The development of radiation-induced toxicities is a multifactorial process. Existing DVH-based prediction models use traditional multivariate statistical methods to combine all the potential risk factors. However, with a dose map based NTCP modelling approach using DL, the combination of radiation dose information with the other potential risk factors is perhaps not as trivial. Multimodality fusion is the next natural step in the process of implementing DL methods for dose map based NTCP modelling. 

As recommended by Huang et al. \cite{Huang2020}, multiple fusion strategies should also be compared and reported in studies combining clinical data with image data with DL methods for radiation-induced toxicity prediction modelling. Our results did not show a statistically significantly difference between fusion strategies, which was likely due to most of the clinical data distributions not being sufficiently different between the ORN and control groups (see Table 1). However, to further investigate this, we aim to repeat this comparison on a larger and more diverse dataset such as the dataset resulting from the PREDMORN (PREDiction models for mandibular OsteoRadioNecrosis in head and neck cancer) multi-institutional study \cite{Humbert-Vidan2022b}.

With regards to technical implementation complexity, late fusion was the least complex strategy. However, unlike early and joint fusion strategies, late fusion does not model the interactions between features from the different data modalities. In traditional DVH-based NTCP modelling, the interactions between dose metrics and clinical variables and the combined effect on the toxicity outcome are modelled via multivariate statistics. Thus, in the context of DL-based NTCP modelling, it is important that these interactions continue to be considered. 
Joint fusion had the highest technical complexity but also resulted in a reduced training time as it only involved one training and hyperparameter optimisation process (the other two fusion strategies consist of two networks each). Moreover, in the joint fusion strategy both the modality-specific and cross-modal patterns are captured during training.

In late fusion, the predicted probabilities from both modality-specific models were directly averaged to calculate the final classification probabilities. This strategy could be further optimised by exploring other methods for combining the outputs, such as a weighted average, which could potentially improve the final predictive performance.

In the early and joint strategies evaluated in our study, the extracted image features (over six hundred) are directly concatenated with the seven clinical features included. DL models are capable of learning complex relationships from large feature spaces or vectors. However, the large feature imbalance between the two modalities could affect the model’s ability to focus on the most discriminative features and less informative features from the dominant modality could (incorrectly) be given more importance. One approach to addressing this feature imbalance could be to assign feature weights based on their importance learned via attention mechanisms \cite{Brauwers2023}. 

The concept of dynamic multimodal fusion \cite{Xue2022,Han2022} has been recently introduced to adaptively fuse multimodal input data during inference. The informativeness is modelled for each modality and feature and used to adjust the importance of the features/modalities in the fusion step. Dynamic fusion could potentially address the feature imbalance issue as well as optimise the fusion strategy based on the data characteristics. 

Finally, this study was developed on a 1:1 matched control-case cohort that did not represent the real-world ORN prevalence. A class-balanced dataset is technically convenient when developing DL classification models. However, the predicted classification probabilities cannot be considered actual ORN risk predictions. Post-processing steps such as probability calibration are necessary prior to clinical implementation of the model to ensure that these are representative of the actual ORN risk.

\section{Conclusion}
Radiation toxicity modelling is rapidly embracing DL methods as the need to move from DVH to spatial dose information is acknowledged. However, how the dose distribution images and clinical data should be combined in a DL pipeline for an optimal NTCP prediction performance has not been fully explored yet. In this study, we aimed to suggest and evaluate pipelines for the early, joint and late fusion strategies and provide a discussion that we hope will inspire future DL-based NTCP modelling studies.

\section{Acknowledgements}
We gratefully acknowledge the support of NVIDIA Corporation with the donation of the Titan Xp GPU used for this research. This work was supported by the Radiation Research Unit at the Cancer Research UK City of London Centre Award [C7893/A28990] and by the Guy’s Cancer Charity via a donation from the Wilson-Olegario foundation and other donations.

\bibliographystyle{unsrt}  
\bibliography{main}  




\end{document}